\documentclass[final]{aipproc}

\layoutstyle{6x9}

\newcommand{\bi}{\begin{itemize}}
\newcommand{\ei}{\end{itemize}}

\newcommand{\be}{\begin{equation}}
\newcommand{\ee}{\end{equation}}
\newcommand{\bea}{\begin{eqnarray}}
\newcommand{\eea}{\end{eqnarray}}

\newcommand{\ldm}{\Delta m_{31}^2}
\newcommand{\sdm}{\Delta m_{21}^2}
\newcommand{\deltacp}{\delta_{\mathrm{CP}}}
\newcommand{\stheta}{\sin^2 2 \theta_{13}}

\newcommand{\ie}{{\it i.e.}}

\newcommand{\eg}{{\it e.g.}}

\newcommand{\Ref}{Ref.}
\newcommand{\Refs}{Refs.}

\newcommand{\Secs}{Secs.}

\newcommand{\Tab}{Table}

\begin{document}

\begin{flushright}
\footnotesize{TUM-HEP-525/03}\\
\end{flushright}

\vspace*{-0.9cm}

\title{Resolving degeneracies for different values of $\mathbf{\theta_{13}}$}

\author{Walter Winter}{
  address={Institut f\"ur theoretische Physik, Physik--Department,
       Technische Universit\"at M\"unchen (TUM),
       James--Franck--Strasse, 85748~Garching bei M{\"u}nchen, Germany}
}

\begin{abstract}
We discuss options to resolve correlations and degeneracies with combinations of future neutrino long-baseline experiments. We  use a logarithmic scale of $\stheta$ as a representation for a systematical classification of the experiments.
\end{abstract}

\maketitle



The most interesting parameters to be determined by future long-baseline experiments are $\theta_{13}$, $\deltacp$, and the mass hierarchy. All of these parameters are suppressed by the size of $\theta_{13}$ itself, which is constrained by the CHOOZ experiment~\cite{Apollonio:2002gd} to below $\stheta \sim 10^{-1}$. From statistics and systematics only, each type of long-baseline experiment has (to a first crude approximation) a characteristic scale of $\theta_{13}$ which it can access with respect to these measurements, \ie, the sensitivity reach in $\theta_{13}$. Unfortunately, the measurements are spoilt by multi-parameter correlations and intrinsic degeneracies, which are the $(\deltacp,\theta_{13})$~\cite{Burguet-Castell:2001ez}, $\mathrm{sgn}(\Delta
m_{31}^2)$~\cite{Minakata:2001qm}, and $(\theta_{23},\pi/2-\theta_{23})$~\cite{Fogli:1996pv} degeneracies leading to an overall ``eight-fold'' degeneracy~\cite{Barger:2001yr}. In this talk, we discuss options to resolve correlations and degeneracies from the point of view of the yet unknown true value of  $\theta_{13}$.
 
We define future long-baseline experiments as neutrino oscillation experiments which are using an artificial neutrino source (\eg , accelerator or reactor) and are sensitive to atmospheric neutrino oscillations, \ie ,  $\ldm L/E = \mathcal{O}(1)$. Compared to a natural neutrino source, such as the atmosphere or the sun, the artificial neutrino source produces a better known neutrino flux. In addition, a near detector is often proposed for a better control of the systematics. Beyond conventional beam experiments, such as K2K~\cite{Nakamura:2001tr}, MINOS~\cite{Paolone:2001am}, and CNGS~\cite{Ereditato:2001an}, future superbeam experiments~\cite{Itow:2001ee,Ayres:2002nm,Beavis:2002ye} are designed to find $\stheta$ down to $\sim 10^{-2}$. Superbeam upgrades\footnote{In this talk, we refer to ``superbeam upgrades'' as superbeams with target powers in the megawatt, and fiducial detector masses in the megaton region, whereas ``superbeams'' refer to the size of the first-generation experiments.}, such as the JHF to Hyper-Kamiokande superbeam~\cite{Itow:2001ee}, could even access $\stheta$ down to  $\sim 10^{-3}$. A useful experiment type below $\stheta \sim 10^{-3}$ (or $\theta_{13} \sim 1^\circ$) is the neutrino factory (see, for example, \Ref~\cite{Apollonio:2002en}). A different type of long-baseline experiment, which also fits our definition, is a reactor experiment with a near and far detector~\cite{Martemyanov:2002td,Minakata:2002jv,Huber:2003pm,Shaevitz:2003ws}, which may find $\stheta$ down to $10^{-2}$ independent of correlations and degeneracies. However, such an experiment does not have sensitivities to $\deltacp$ or the mass hierarchy at all.


There are essentially six different impact factors, which determine the performance of future long-baseline experiments (for a more detailed discussion, see \Secs~3 and~5 of \Ref~\cite{Huber:2002mx}):
\begin{description}
\item[1. Statistical errors] describe the experiment performance from statistics only.
\item[2. Systematics] makes the statistical errors somewhat larger and is determined by the experiment itself.
\item[3. Correlations] are connected degenerate solutions (at the chosen confidence level). The measurement error of the quantity of interest is usually obtained as the projection of the $n$-dimensional connected manifold onto the respective axis. It can even be orders of magnitude larger than the original error.
\item[4. Degeneracies] are disconnected degenerate solutions (at the chosen confidence level). Their treatment in the results depends on the definition of the quantity of interest. 
\item[5. External input] can partially resolve correlations and degeneracies. For the long-baseline experiments, the external input usually includes the solar measurements and the knowledge about the matter density, whereas the atmospheric oscillation parameters are normally obtained by the disappearance channels. 
\item[6. The true parameter values] are only known with a certain precision before the experiments are built. It can be shown that for future long-baseline experiments (depending on the measurement) especially the true values of $\sdm$ (within the KamLAND-allowed range), $\ldm$, $\deltacp$, and $\theta_{13}$ itself determine the performance.  
\end{description}
These six different impact factors can be arranged in three different groups:
\begin{description}
\item[1. (Statistics) and 2. (Systematics)] are determined by the R\&D of the experiment.
\item[3. (Correlations) and 4. (Degeneracies)] are reducible by clever choices of baseline, energy, and combinations of experiments. 
\item[5. (External input) and 6. (True parameter values)] are not controllable by the considered experiment at all. 
\end{description}
Experimental results often include statistics and systematics, but not (or only partially) correlations and degeneracies. From the theoretical point of view, however, especially the correlations and degeneracies are relevant for the optimization of experiments. Thus, it is necessary that a quoted measurement error be clearly defined in order to be comparable to that of other analyses. In this talk, we only refer to the second group, \ie, the reduction of correlations and degeneracies by the choices of baseline, energy, and combinations of experiments.

As far as correlations and degeneracies are concerned, there is an important difference between the analysis of existing and future experiments. An existing experiment would obtain one or more regions in parameter space fitting the data, such as the formerly allowed solar solutions. The purpose of the analysis of future experiments is, however, to minimize the extension (\ie, correlations) and number (\ie, degeneracies) of the disconnected solutions {\em before} the experiments are going to be built. In addition, the risk with respect to the yet unknown true parameter values within their allowed regions should be minimized. Thus, condensing the information as function of the most relevant parameters by a reasonable inclusion of correlations and degeneracies in the analysis is crucial for the optimization of future experiments.


\begin{table}[t!]
\begin{tabular}{p{2cm}p{2.2cm}p{2.2cm}p{2.2cm}p{2.5cm}}
\hline
$\mathbf{\sin^2 2 \theta_{13}}$ &$\mathbf{\sim 10^{-1}-10^{-2}}$ & $\mathbf{\sim 10^{-2}-10^{-3}}$  & $\mathbf{\sim 10^{-3}-10^{-4}}$ &  $\mathbf{ < 10^{-4}}$ \\[0.15cm]
$\theta_{13}$ [Degrees] & $\sim 9^\circ - 3^\circ$ & $\sim 3^\circ - 1^\circ$ & $\sim 1^\circ - 0.3^\circ$ & $< 0.3^\circ$ \\[0.15cm] 
Timescale??? & $2000-2010$ & $2010-2025$ & $2025-2040$ & $>2040$ \\
\hline
 \bf{Sensitive} \newline \bf{experiments:} & \mbox{- Conventional} \newline \mbox{~~beams (partially)}   \newline
- Reactor \newline \mbox{~~experiments}  \newline  
\mbox{- First-generation} \newline \mbox{~~superbeams} & 

\mbox{- Superbeam} \newline \mbox{~~upgrades} \newline 
\mbox{- Neutrino} \newline \mbox{~~factories} \newline
- Reactor \newline \mbox{~~upgrades?} & 
\mbox{- Neutrino} \newline \mbox{~~factories}
& 
\mbox{-  Neutrino} \newline \mbox{~~factories?} \newline 
\mbox{- Theoretical} \mbox{~~reason for} \newline \mbox{~~$\sin^2 2 \theta_{13} \equiv 0$?} \\
\hline
\end{tabular}
\caption{Logarithmic scale of the true value of $\stheta$, the corresponding values in degrees, a possible timescale, and the experiments, which are sensitive in the respective intervals. Note that the interval limits are only a crude approximation, and that in some cases the experiments do not entirely cover the whole intervals.}
\label{tab:scales} 
\end{table}

In \Tab~\ref{tab:scales}, a logarithmic scale of $\stheta$ is shown together with the relevant future long-baseline experiments for the considered ranges in the respective columns. Note that the ranges are only crude approximations for the $\stheta$, CP violation, and mass hierarchy sensitivity reaches under optimal conditions, \ie, statistics and systematics only.
\Tab~\ref{tab:scales} implies that it only makes sense to combine experiments with similar capabilities, which are reflected by $\stheta$ as well as the timescale. Thus, by combining experiments from different columns in \Tab~\ref{tab:scales} the results are normally dominated by the statistics of the better experiment.  Thus, one could also say that the true value of $\stheta$ ``selects'' the experiment types sensitive to it, which can be read off \Tab~\ref{tab:scales}. Of course, this picture is quite simple and does not take into account continuously adjustable luminosities of the different experiment types. However, it is quite illustrative to classify possible experiment combinations to resolve correlations and degeneracies, which are so far:
\begin{description}
\item[$\mathbf{\stheta \sim 10^{-1} - 10^{-2}}$:] In this range, one can combine two first-generation superbeams~\cite{Wang:2001ys,Whisnant:2002fx,Barger:2002rr,Huber:2002rs,Minakata:2003ca} or one or two superbeams with a reactor experiment~\cite{Minakata:2002jv,Huber:2003pm} to resolve the degeneracies.
\item[$\mathbf{\stheta \sim 10^{-2} - 10^{-3}}$:] One may combine two superbeam upgrades~\cite{Barger:2002xk}, a superbeam upgrade with a neutrino factory~\cite{Burguet-Castell:2002qx}, the ``golden'' $\nu_e \leftrightarrow \nu_\mu$ appearance channel at a neutrino factory or superbeam upgrade with the ``silver'' channel ($\nu_\tau$ detection) at a neutrino factory~\cite{Donini:2002rm,Autiero:2003fu}, or operate a superbeam upgrade at the ``magic baseline''~\cite{Asratyan:2003dp} (see also below).
\item[$\mathbf{\stheta \sim 10^{-3} - 10^{-4}}$:] Since so far only neutrino factories have been demonstrated to operate efficiently in this range, one can only combine two neutrino factory baselines, which are naturally obtained from one neutrino factory. It has been realized in numerical (see, for example, \Refs~\cite{Burguet-Castell:2001ez,Petcov:2001vu}) and analytical~\cite{Barger:2001yr,Lipari:1999wy} analyses that at a baseline $L_{\mathrm{magic}} \sim 7 \, 300 \, \mathrm{km}$ all correlations and degeneracies involving the CP phase vanish independent of the energy and the oscillation parameters. It has therefore been  called ``magic baseline''~\cite{Huber:2002uy,Lindner:2002pj}. In \Ref~\cite{Huber:2003ak}, it has finally been  demonstrated that the $\stheta$, (maximal) CP violation, and mass hierarchy sensitivities are all good in the considered range $\stheta \sim 10^{-3} - 10^{-4}$ for the combination of the two baselines $L_1 \sim 3 \, 000 \, \mathrm{km}$ and $L_2 = L_{\mathrm{magic}}$.
\end{description}


In summary, we have discussed the combination of experiments to resolve degeneracies based upon the true value of $\stheta$, which ``selects'' the experiments sensitive to the quantities of interest. These sensitive experiments can then be combined to resolve correlations and degeneracies. The discussion in this talk is different from a strategical one, since the strategy depends on when $\stheta$ is found and the information and experiments available at that time. Right now, of course, the most interesting range is $\stheta \sim 10^{-1} - 10^{-2}$, which means that later decisions will depends on the results within this range and new technologies could still change the following options. Finally, we have discussed a logarithmic scale of $\stheta$, which is not necessarily an appropriate representation for all purposes. For example, a linear scale may be more plausible from linear mass models in flavor space. 
 
I would like to thank P.~Huber and M.~Lindner
for useful discussions and comments. This work was supported by the ``Studienstiftung des deutschen Volkes'' and the ``SFB 375 f{\"u}r Astro-Teilchenphysik der DFG''.

\renewcommand{\refname}{}

\end{document}